\documentclass[preprint,preprintnumbers,amsmath,amssymb]{revtex4}
\usepackage{graphicx}
\usepackage{bm}

\usepackage{mhchem}
\usepackage{bm}
\usepackage{amsmath}
\usepackage{amsfonts}
\usepackage{amssymb}

\usepackage{breqn}
\usepackage{subfigure}
\usepackage{balance}

\usepackage{times,mathptmx}

\newcommand{\id}{{\bm 1}}

\newcommand{\be}{\begin{equation}}
\newcommand{\ee}{\end{equation}}
\newcommand{\bea}{\begin{eqnarray}}
\newcommand{\eea}{\end{eqnarray}}

\begin{document}
\title{Internal dynamics and activated processes in Soft-Glassy materials}
\author{R.Benzi$^{1}$, M. Sbragaglia$^{1}$, A. Scagliarini$^{1}$, P. Perlekar$^{2}$, M. Bernaschi$^{3}$, S. Succi$^{3}$ and F. Toschi$^{4}$\\
$^{1}$ Department of Physics and  INFN, University of ``Tor Vergata'', Via della Ricerca Scientifica 1, 00133 Rome, Italy\\
$^{2}$ Istituto per le Applicazioni del Calcolo CNR, Viale del Policlinico 137, 00161 Roma, Italy\\
$^{3}$ TIFR Centre for Interdisciplinary Sciences, 21 Brundavan Colony, Narsingi, Hyderabad 500075, India\\
$^{4}$ Department of Physics and Department of Mathematics and Computer Science and J.M. Burgerscentrum, Eindhoven University of Technology, 5600 MB Eindhoven\\
}

\begin{abstract} 
Plastic rearrangements play a crucial role in the characterization of soft-glassy materials, such as emulsions and foams. Based on numerical simulations of soft-glassy systems, we study the dynamics of plastic rearrangements at the hydrodynamic scales where thermal fluctuations can be neglected. Plastic rearrangements require an energy input, which can be either provided by external sources, or made available through time evolution in the coarsening dynamics, in which the total interfacial area decreases as a consequence of the slow evolution of the dispersed phase from smaller to large droplets/bubbles. We first demonstrate that our hydrodynamic model can quantitatively reproduce such coarsening dynamics. Then, considering periodically oscillating strains, we characterize the number of plastic rearrangements as a function of the external energy-supply, and show that they can be regarded as activated processes induced by a suitable "noise" effect. Here we use the word noise in a broad sense, referring to the internal non-equilibrium dynamics triggered by spatial random heterogeneities and coarsening. Finally, by exploring the interplay between the internal characteristic time-scale of the coarsening dynamics and the external time-scale associated with the imposed oscillating strain, we show that the system exhibits the phenomenon of stochastic resonance, thereby providing further credit to the mechanical activation scenario.
\end{abstract}

\maketitle

Foams and emulsions are dispersions of a fluid phase in a liquid phase, with a suitable stabilization mechanism for the interface. The dispersed phase is constituted by gas bubbles in foams and liquid droplets in emulsions. The dynamics of such complex fluids is usually characterized by a large interface area and a relatively large packing fraction of the dispersed phases, giving rise to a genuinely new branch of non-equilibrium thermodynamics which has witnessed a major upsurge of interest in the last decades, mostly in connection with soft-matter research \cite{Larson,Coussot,FOAMrev}. As a general rule, a process called {\it coarsening} takes place to minimize interface area: the dispersed phase diffuses through the continuous one to reduce the total area of the interface. On average, large domains expand  at the expenses of smaller ones that shrink and eventually disappear \cite{Lambert10}. For dilute systems, domains are separated spheres, uniquely characterized by their radii. Their coarsening dynamics is known as ``Ostwald ripening'', or ``LSW'' after Lifshitz, Slyosov, and Wagner \cite{Ostwald,Lifschitz,Wagner}.  In the opposite limit of a dry diphasic system, the continuous matrix occupies a much  smaller volume than the dispersed phase and domains resemble closely packed polyhedra.  In this case, diffusion takes place through thin walls of roughly constant thickness (Plateau domains), thus triggering a different coarsening mechanism.  In two dimensions, for example, the growth rate of each domain is proportional  to its number of neighbors, not on its size,  in agreement with  the celebrated Von-Neumann's law \cite{VonNeumann}. In such a jammed state, the discrete objects are locked into a mechanically  stable packing configuration, where neighbors cannot change just due to thermal motion. This happens because the single constituents are macroscopic in size and the corresponding energy barriers for droplets/bubbles rearrangements are orders of magnitude larger than $k_BT$. The system thus exhibits a yield stress $\sigma_Y$, below which it deforms elastically instead of flowing.  Above the yield stress threshold, the system flows with a rheo-thinning character, and  flow takes place through a sequence of  irreversible plastic rearrangements \cite{Goyon,Bocquet09,Mansard11,Jop12,Barrat13}. In emulsions and foams, plastic rearrangements are identified with T1 topological events, in which one edge of a given droplet/bubble collapses to zero and neighbour droplets/bubbles switching happens. For such rearrangements to occur, i.e., to expose a liquid-like character, an input  of energy is required. This mechanical energy-supply can result from an external source, namely stress or strain imposed to the system. However, this is not a necessary condition, as the input may also be  supplied internally, i.e. via dynamic rearrangements induced by the coarsening. This is the case, for example, of the coarsening-induced bubble dynamics in real  foams, experimentally studied in a series of papers (see \cite{Durian08,Durian13} and references therein). In this paper, we aim at investigating whether plastic rearrangements can be regarded as "activated processes", i.e. processes induced by a suitable noise effect. The word "noise" is used here in a rather broad sense without any reference to a thermodynamic interpretation. To qualify the meaning of "activated process", we think of the classical picture originally proposed by Kramer \cite{Kramer40,SR1,Barratrecent}, where an overdamped particle can escape a potential well due to the noise effect. Activated processes are rare events and the time needed for the escape to occur is a random variable with an exponential distribution. Moreover the average escape time depends exponentially on the potential barrier, hence one can predict the rate of activated processes as a function of an external driving force. If the force is periodic, the system may exhibit a behaviour known as {\it stochastic resonance} (SR) \cite{SR1,SR2}, i.e. activated processes are in phase with the external driving force for a well defined frequency.\\
To the purpose of this paper, we explore the properties of plastic rearrangements at changing an imposed deformation under a periodically oscillating strain. The number of plastic events per cycle is shown to depend exponentially on the external energy-supply, thus supporting their interpretation as activated processes. It is also found that the statistical dynamics of plastic rearrangements is compatible with a SR mechanism \cite{SR1,SR2}, which provides a specific scenario for the mechanical activation process, where the source of ``noise'' is the internal non-equilibrium dynamics triggered by random spatial heterogeneities in the interface geometry. We also posit that such level of noise is regulated by the polydispersity of the dispersed phase, which gives rise to a random distribution of free-energy barriers within the diphasic system. The idea of an activated escape from free-energy barriers due to effective noise is not new and it has been vigorously pursued within the so-called SGR (Soft-Glassy Rheology) model proposed years back by  Sollich {\em et al.} \cite{SGR1,SGR2,SGR3,HebraudLequeux,Bocquet09}. The SGR builds on Bouchaud's trap model \cite{Bouchaud_trap}, based on the assumption that temperature alone is unable to achieve complete structural relaxation. Hence, an effective temperature is introduced in the system as the relevant source for plastic rearrangements to occur. Similarly, other authors pointed to the central role played by the concept of an effective temperature characterizing the state of configurational disorder in amorphous systems \cite{Ono02,Ilg07,Langer04,Bouchbinder07a,Bouchbinder07b,Langer07}. \\
The numerical results presented in this paper are obtained with a mesoscopic Lattice Boltzmann (LB) method \cite{CHEM09,EPL10,EPL13}, which allows the simulation of droplets and their interface motion under different load conditions. The essential features of the model are recalled in section \ref{sec:numericalmodel}. In section \ref{sec:aging}, we explore the coarsening properties of dry diphasic systems and show that our model proves capable of reproducing the Von-Neumann topological evolution rules \cite{VonNeumann}. In such a jammed state, we extract the characteristic internal time-scale of the system, taken as the average rest time between successive rearrangements\cite{Durian08,FOAM}. Section \ref{sec:response} is fully devoted to the characterization of plastic rearrangements in presence of an externally applied oscillating strain. We will finally explore the interplay between the internal characteristic time-scale of the coarsening dynamics and the external  time-scale associated with the imposed periodic oscillating strain, showing that the system exhibits SR behaviour. Summary and outlook follow in Section \ref{sec:conclusions}.

\section{The Lattice Kinetic Model}\label{sec:numericalmodel}

\begin{figure}[t!]
\includegraphics[width=0.48\textwidth]{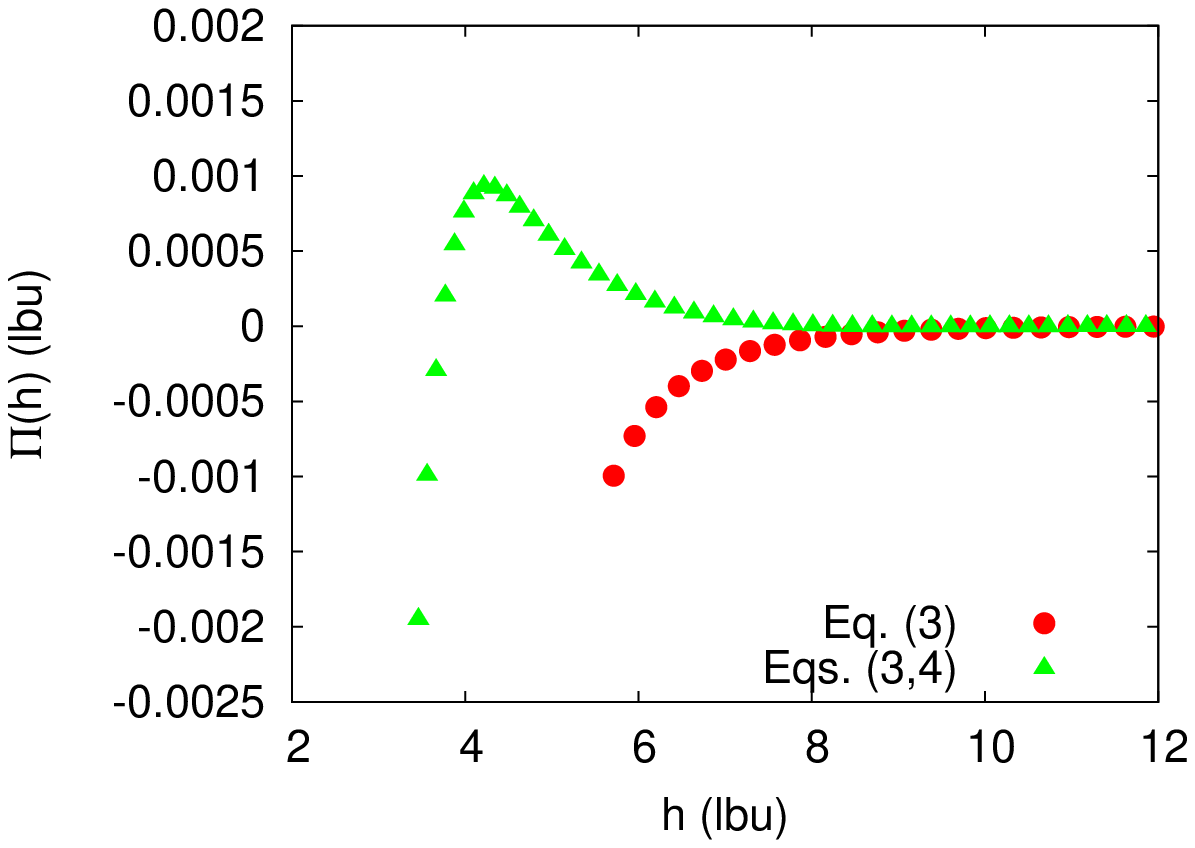}\\
\subfigure[\,\, Eq. (3), t=t$_0$]{\includegraphics[width=0.155\textwidth]{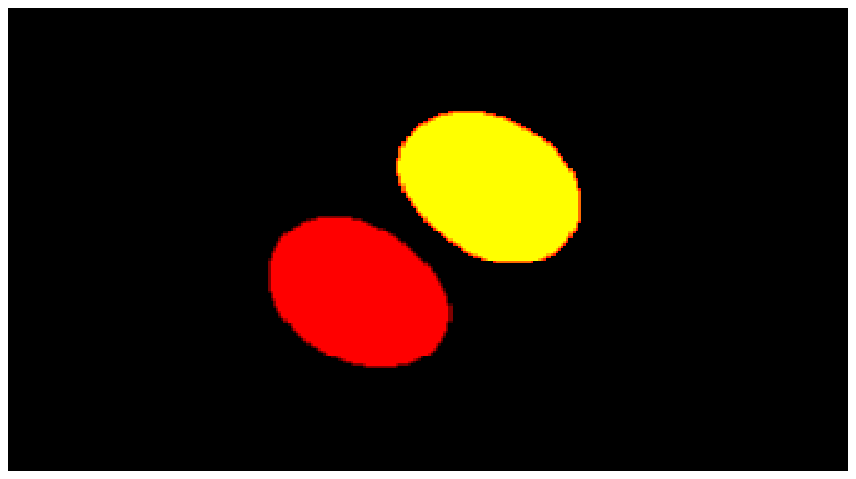}}
\subfigure[\,\, Eq. (3), t=t$_{1}$]{\includegraphics[width=0.155\textwidth]{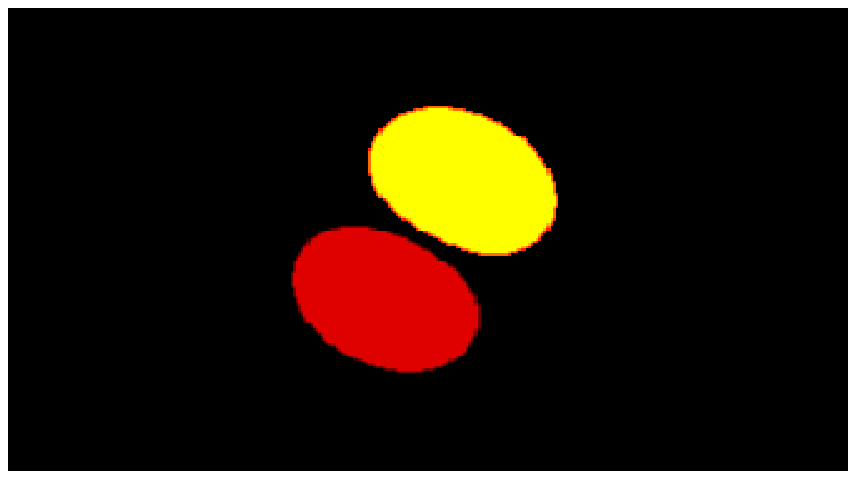}}
\subfigure[\,\, Eq. (3), t=t$_{2}$]{\includegraphics[width=0.155\textwidth]{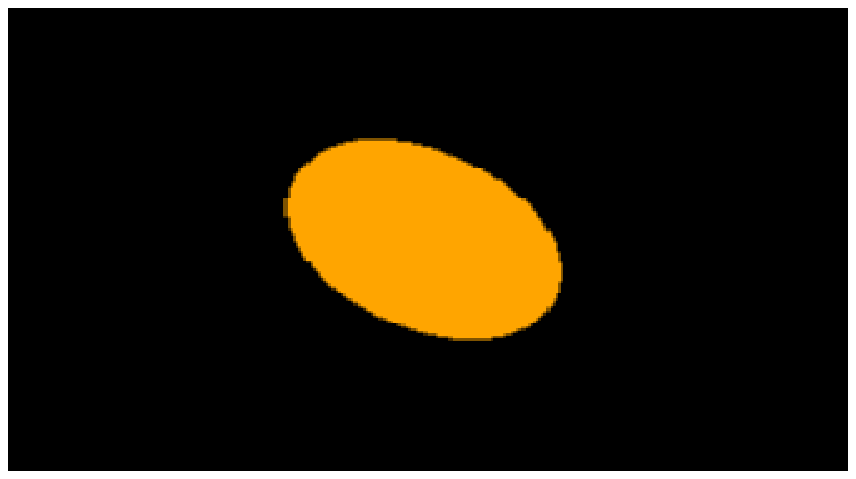}}\\
\subfigure[\,\, Eq. (3,4), t=t$_0$]{\includegraphics[width=0.155\textwidth]{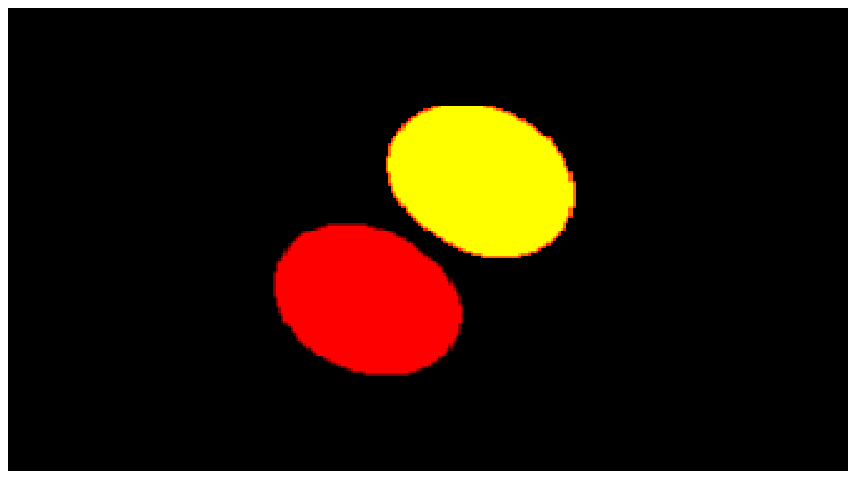}}
\subfigure[\,\, Eqs. (3,4), t=t$_1$]{\includegraphics[width=0.155\textwidth]{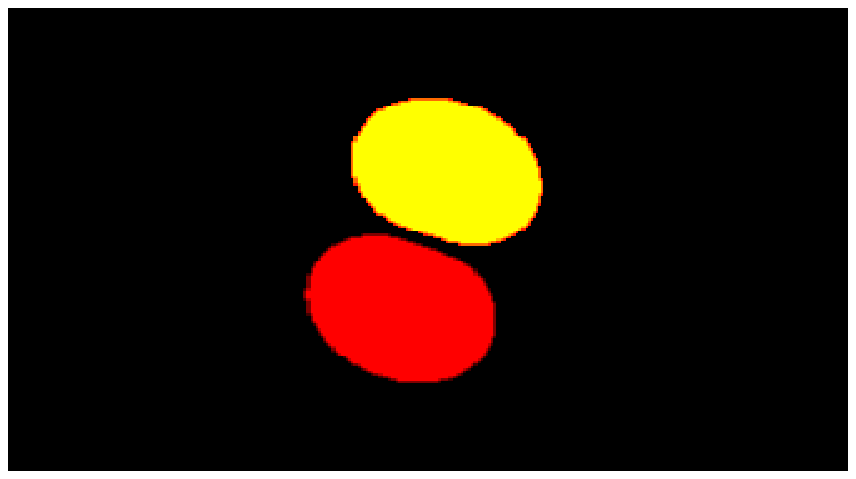}}
\subfigure[\,\, Eqs. (3,4), t=t$_2$]{\includegraphics[width=0.155\textwidth]{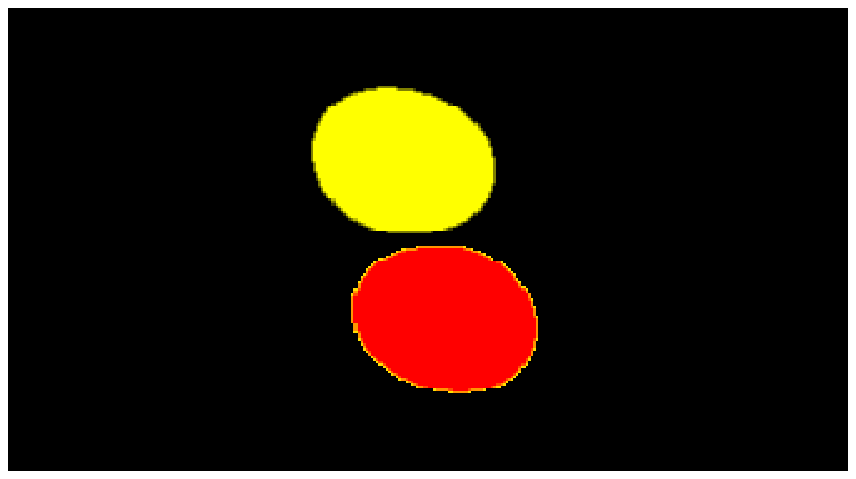}}
\caption{Emergence of disjoining pressure $\Pi(h)$ in a thin {\it film} characterized by two non-ideal interfaces, a distance $h$ apart. The disjoining pressure is measured (see text for details) for the LB model \eqref{LB} with phase separating interactions \eqref{Phase} supplemented with competing interactions \eqref{NNandNNN}. Competing interactions \eqref{NNandNNN} are instrumental to achieve a positive disjoining pressure stabilizing thin films between neighboring droplets. To test such stabilization, we report some snapshots out of a very simple experiment: two droplets are brought close to each other progressively in time ($t_0<t_1<t_2$) by an imposed shear flow. Different colors refer to different droplets. In absence of competing interactions \eqref{NNandNNN} (Panels (a)-(c)), the two droplets merge into a single droplet (Panel (c)). This is quantitatively attributed to the presence of a negative disjoining pressure (filled circles in the main panel). The natural tendency of the interface to minimize its area via merger events is ``frustrated'' by the presence of competing interactions \eqref{NNandNNN} (Panels (d)-(f)) which allow to achieve a positive disjoining pressure (filled triangles in the main panel). Simulation parameters are reported in the text.}
\label{fig:disjoining}
\end{figure}

In recent years we developed a mesoscopic approach to the description of soft-glassy rheology, based on a lattice  version of Boltzmann (LB) kinetic equation with competing short-range attraction and mid-range repulsion (see \cite{CHEM09,EPL10,SOFT12,EPL13} and supplementary material). The lattice kinetic model has been shown to reproduce, with a pretty good quantitative agreement, many distinctive features of soft-glassy materials, such as ageing \cite{CHEM09} and non-linear cooperative rheology \cite{EPL10}. Specifically, it allows, at an affordable computational cost, to simulate  a collection of closely packed droplets with variable polydispersity and packing fraction, under different load conditions \cite{EPL13}. Since our lattice kinetic model has been already described in several previous works \cite{CHEM09,EPL10}, we recall here just its basic features. We start from a mesoscopic lattice Boltzmann model for non-ideal binary fluids, which combines a small positive surface tension, promoting highly complex interfaces, with a positive disjoining pressure, inhibiting interface coalescence. The mesoscopic kinetic model considers two fluids $A$ and $B$, each described by a {\it discrete} kinetic distribution function $f_{\zeta i}({\bm r},{\bm c}_i;t)$, measuring the probability of finding a particle of fluid $\zeta =A,B$ at position ${\bm r}$ and time $t$, with discrete velocity ${\bm c}_i$, where the index $i$ runs over the nearest and next-to-nearest neighbors of ${\bm r}$ in a regular two-dimensional  lattice \cite{BCS,CHEM09}. In other words, the mesoscale particle represents all molecules contained in a unit cell of the lattice. The distribution functions evolve in time under the effect of free-streaming and local two-body collisions, described, for both fluids ($\zeta=A,B$), by a relaxation towards a local equilibrium ($f_{\zeta i}^{(eq)}$) with a characteristic time-scale $\tau_{LB}$:
\begin{dmath}\label{LB}
f_{\zeta i}({\bm r}+{\bm c}_i,{\bm c}_i;t+1) -f_{\zeta i}({\bm r},{\bm c}_i;t)  = -\frac{1}{\tau_{LB}} \left(f_{\zeta i}-f_{\zeta i}^{(eq)} \right)({\bm r},{\bm c}_i;t)+F_{\zeta i}({\bm r},{\bm c}_i;t).
\end{dmath}
The equilibrium distribution is given by
\be
f_{\zeta  i}^{(eq)}=w_i \rho_{\zeta} \left[1+\frac{{\bm u} \cdot {\bm c}_i}{c_s^2}+\frac{{\bm u}{\bm u}:({\bm c}_i{\bm c}_i-c_s^2 \id)}{2 c_s^4} \right]
\ee
with $w_i$ a set of weights known a priori through the choice of the quadrature \cite{Sbragaglia07,Shan06}. The model gives direct access to the hydrodynamical variables: coarse grained hydrodynamical densities are indeed defined for both species $\rho_{\zeta }=\sum_i f_{\zeta i}$ as well as a global momentum for the whole binary mixture ${\bm j}=\rho {\bm u}=\sum_{\zeta , i} f_{\zeta i} {\bm c}_i$, with $\rho=\sum_{\zeta} \rho_{\zeta}$. The term $F_{\zeta i}({\bm r},{\bm c}_i;t)$ is just the $i$-th projection of  the total internal force which includes a variety of interparticle forces. First, a repulsive ($r$) force with strength parameter ${\cal G}_{AB}$ between the two fluids
\begin{equation}\label{Phase}
{\bm F}^{(r)}_\zeta ({\bm r})=-{\cal G}_{AB} \rho_{\zeta }({\bm r}) \sum_{i, \zeta ' \neq \zeta } w_i \rho_{\zeta '}({\bm r}+{\bm c}_i){\bm c}_i
\end{equation}
is responsible for phase separation \cite{CHEM09}. Interactions \eqref{Phase} are nothing but a lattice transcription of mean-field models for phase separation (see \cite{Bastea} and references therein). They can produce stable non-ideal interfaces with a positive surface tension \cite{SC1,CHEM09}. However, they give rise {\it only} to negative disjoining pressures, i.e. ``thin'' films between neighboring droplets cannot be stabilized against rupture (see Fig. \ref{fig:disjoining}). To provide an energy barrier against the rupture of such thin films, we introduce competing interactions encoding a mechanism of {\it frustration} ($F$) for phase separation \cite{Seul95}. In particular, we model short range (nearest neighbor, NN) self-attraction, controlled by strength parameters ${\cal G}_{AA,1} <0$, ${\cal G}_{BB,1} <0$, and ``long-range'' (next to nearest neighbor, NNN) self-repulsion, governed by strength parameters ${\cal G}_{AA,2} >0$, ${\cal G}_{BB,2} >0$
\begin{dmath}\label{NNandNNN}
{\bm F}^{(F)}_\zeta ({\bm r})=-{\cal G}_{\zeta \zeta ,1} \psi_{\zeta }({\bm r}) \sum_{i \in NN} w_i \psi_{\zeta }({\bm r}+{\bm c}_i){\bm c}_i -{\cal G}_{\zeta \zeta ,2} \psi_{\zeta }({\bm r}) \sum_{i \in NNN} w_i \psi_{\zeta }({\bm r}+{\bm c}_i){\bm c}_i
\end{dmath}
with $\psi_{\zeta }({\bm r})=\psi_{\zeta }[\rho({\bm r})]$ a suitable pseudo-potential function \cite{SC1,SbragagliaShan11}. The pseudo-potential is taken in the form originally suggested by \citet{SC1}
\begin{equation}
\label{PSI}
\psi_{\zeta}[\rho_{\zeta}({\bm r})]= \rho_{0} (1-e^{-\rho_{\zeta}({\bm r})/\rho_{0}}).
\end{equation}
The parameter $\rho_{0}$ marks the density value above which non-ideal effects come into play. The prefactor $\rho_{0}$ in (\ref{PSI}) is used to ensure that for small densities the pseudopotential is linear in the density $\rho_{\zeta}$. In all the simulations presented in this paper, we fix the reference density to be $\rho_0=0.83$ lbu (LB units), the relaxation time $\tau_{LB}=1.0$ lbu in \eqref{LB} and we will change the magnitude of the various  ${\cal G}$'s in \eqref{Phase}-\eqref{NNandNNN} to explore different physical scenarios. The disjoining pressure $\Pi(h)$ is computed for a thin {\it film} with two non-ideal interfaces separated by the distance $h$. Following Bergeron \cite{Bergeron}, we write the relation for the corresponding tensions starting from the Gibbs-Duhem relation:
\be
\begin{split}
\Gamma_f(h)=2 \Gamma+\int^{\Pi(h)}_{\Pi(h=\infty)} h \, d \Pi
\end{split}
\ee
where $\Gamma$ is the bulk value of the surface tension and $\Gamma_f$ is the overall film tension, whose expression is known in terms of the mismatch between the normal and tangential components of the pressure tensor \cite{Toshev,Derjaguin}, $\Gamma_f=\int_{-\infty}^{+\infty} (P_N-P_T(x)) dx$,  where, in our model, $P_N-P_T(x)=p_s(x)$ can be computed analytically \cite{Shan08}. Knowledge of the relation $s(h)=\Gamma_f(h)-2 \Gamma$ makes it straightforward to compute the disjoining pressure: a simple differentiation of $s(h)$ permits  to compute the first derivative of the disjoining pressure, $\frac{d s(h)}{d h}=h \frac{d \Pi}{d h}$.  This information, supplemented with the boundary condition $\Pi(h \rightarrow \infty) = 0$, allows to completely determine the disjoining pressure of the film. The emergence of the disjoining pressure and its effect on the stabilization of thin films between neighboring droplets is illustrated in Fig. \ref{fig:disjoining}, where we report details of a very simple numerical experiment: the computational domain is a rectangular box of size $L_x \times L_z=2 L_z \times L_z $ ($x$ is the stream-flow direction) covered by $N_x \times N_z =400 \times 200$ lattice sites. The boundary conditions are those of a steady velocity at the upper and lower boundaries $\pm U_W=\pm 0.02 $ lbu, whereas periodic boundary conditions are applied in the stream-flow direction. Two droplets with equal radius ($R=40$ lbu) are placed slightly below and above the half-plane of the computational domain ($z=L_z/2$) and brought one close to the other by the imposed flow. One numerical simulation is performed solely with the phase separating interactions \eqref{Phase} with strength parameter ${\cal G}_{AB}=0.586$ lbu, corresponding to bulk densities $\rho_A=1.2$ ($0.12$) lbu and $\rho_B=0.12$ ($1.2$) lbu in the dispersed (continuous) phase. As we can visually appreciate from Panels (a)-(c), when coming close to each other, the two droplets merge into a single droplet. This is quantitatively attributed to the presence of a negative disjoining pressure (see main panel of Fig. \ref{fig:disjoining}, filled circles)  which triggers the rupture of the thin film (roughly $h \approx 8-10$ lattice points) between the two droplets. When competing interactions \eqref{NNandNNN} are added to the phase separating interactions \eqref{Phase}, the disjoining pressure may be changed. Indeed, the filled triangles reported in the main panel of Fig. \ref{fig:disjoining} refer to a case with phase separating interactions \eqref{Phase} with strength parameter ${\cal G}_{AB}=0.586$ lbu, supplemented with competing interactions with strength parameters ${\cal G}_{AA,1}={\cal G}_{BB,1}=-8.0$ lbu and ${\cal G}_{AA,2}={\cal G}_{BB,2}=7.1$ lbu. The disjoining pressure is now positive, thus promoting the emergence of an energy barrier against coalescence, which is adjustable by tuning the competing interactions strengths. This provides a stabilization mechanism for the thin film and the two droplets can ``slide'' one on the other without merging (see panels (d)-(f)). We remark that the computation of the disjoining pressure, as previously discussed, allows us to understand the role of the competing interactions \eqref{NNandNNN} in the formation of a stable interface: by changing the parameters we can either increase or decrease the peak in $\Pi(h)$ reported in Fig. \ref{fig:disjoining}. Therefore, by increasing such peak, we can argue that there would be also a slowing down of the diffusion processes (see Section \ref{sec:aging}) and, consequently, longer time scales for coarsening to occur. This will be quantitatively explored in the next section. \\
Before closing this section, we wish to stress that the micro-mechanics of the model, Eqs. \eqref{Phase}-\eqref{NNandNNN}, is not meant to mimic ``distinct'' physico-chemical details of a real system, but rather to model a ``generic'' soft-glassy system with {\it non-ideal fluid behaviour} (e.g., non-ideal equation of state, phase separation), {\it interfacial phenomena} (e.g., surface tension, disjoining pressure) and {\it hydrodynamics}.



\section{Aging and Internal Rearrangements}\label{sec:aging}

\begin{figure}[t!]
\subfigure[\,\, Eq. (3), t=t$_0$]{\includegraphics[width=0.155\textwidth]{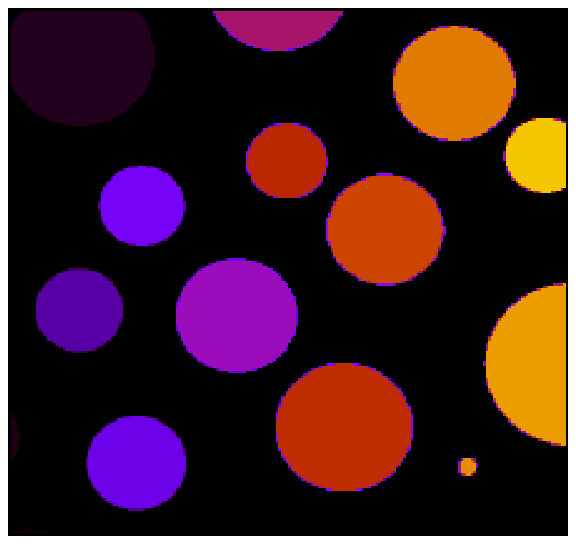}}
\subfigure[\,\, Eq. (3), t=t$_0$+$\Delta$]{\includegraphics[width=0.155\textwidth]{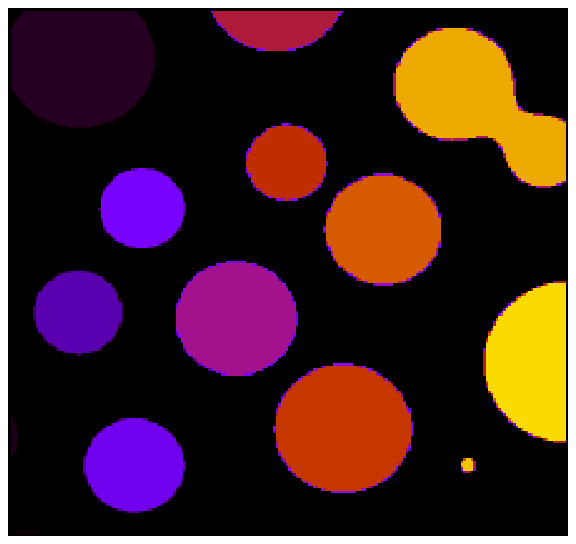}}
\subfigure[\,\, Eq. (3), t=t$_0$+$2\Delta$]{\includegraphics[width=0.155\textwidth]{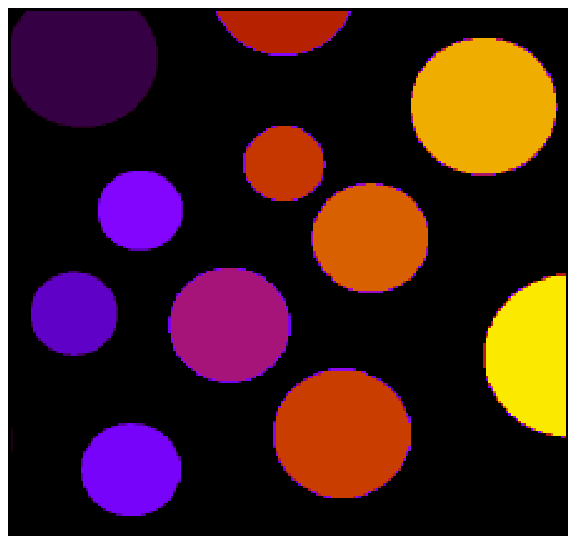}}\\
\subfigure[\,\, Eq. (3,4), t=t$_0$]{\includegraphics[width=0.155\textwidth]{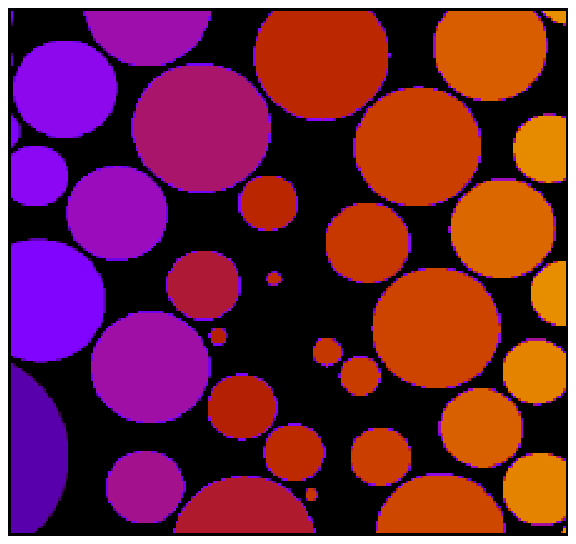}}
\subfigure[\,\, Eq. (3,4), t=t$_0$+$4\Delta$]{\includegraphics[width=0.155\textwidth]{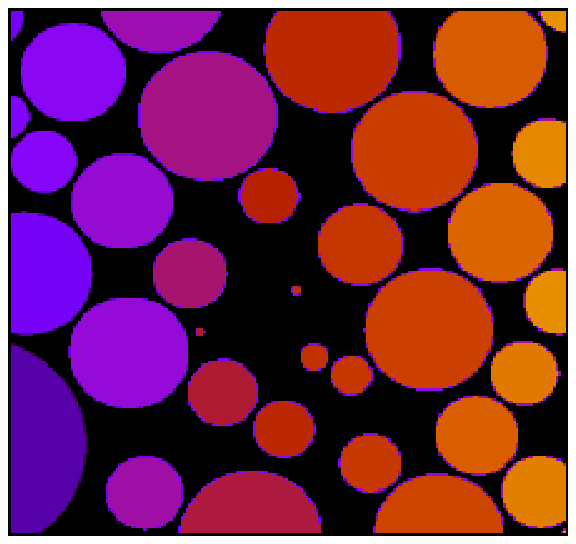}}
\subfigure[\,\, Eq. (3,4), t=t$_0$+$7\Delta$]{\includegraphics[width=0.155\textwidth]{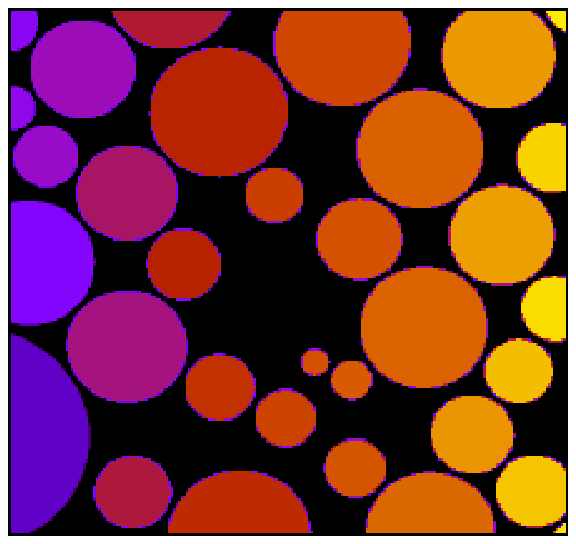}}\\
\includegraphics[width=0.48\textwidth]{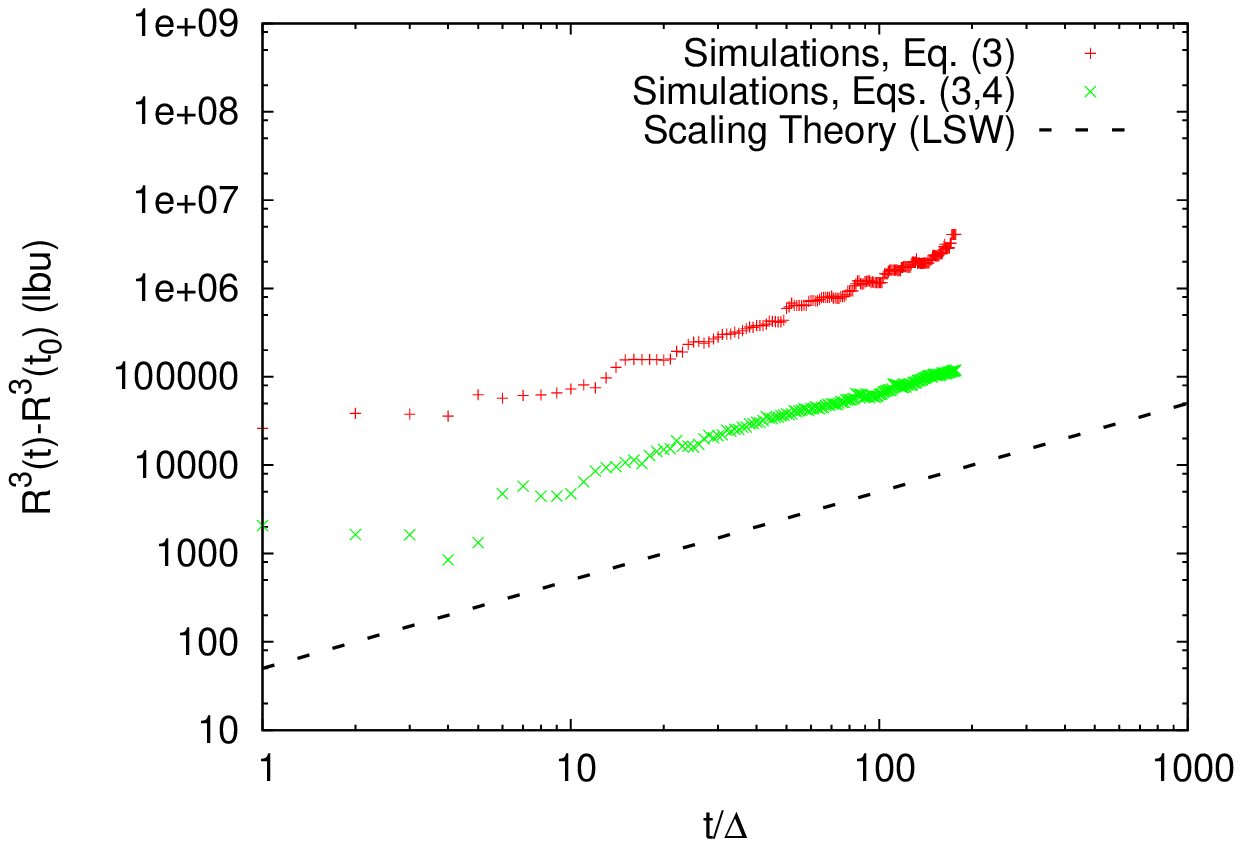}
\caption{Coarsening dynamics for the LB model \eqref{LB} with phase separating interactions \eqref{Phase} and competing interactions \eqref{NNandNNN} in the {\it dilute} limit. Different colors refer to different droplets. The standard LSW \cite{Ostwald,Lifschitz,Wagner} scenario is realized with the phase separating interactions \eqref{Phase} only, allowing for a coarsening dynamics determined by both coalescence and mass diffusion ($+$ and Panels (a)-(c), reporting snapshots of a sub-region of the system). The scaling law of the average droplet radius, $R(t)$, quantitatively matches the prediction of the LSW theory, $R^3(t) \sim t$ (dashed line). When competing interactions  \eqref{NNandNNN} supplement phase separating interactions \eqref{Phase}, they switch-off coalescence and slow down the coarsening dynamics with a much smaller prefactor in front of the LSW scaling law ($\times$ and Panels (d)-(f)). Simulation parameters are reported in the text. $\Delta=3 \times 10^4$ lbu is a characteristic time lapse. }
\label{fig:coarsening-dilute}
\end{figure}

In this section we quantify the role of frustration in the coarsening dynamics of both {\it dilute} and {\it dry} systems, to show that the model actually reproduces well known features of coarsening in the aforementioned limits \cite{Ostwald,Lifschitz,Wagner,VonNeumann}. More importantly, the analysis of coarsening in the dry limit will be instrumental to characterize the internal time-scale of the system in the jammed state \cite{Durian08,FOAM}. It also provides some hint about the meaning of the noise acting in the system.\\
In presence of a collection of droplets with a heterogeneous distribution of radii, even the inhibition of coalescence does not prevent macroscopic phase separation of the droplets. This is because the Ostwald ripening (or LSW \cite{Ostwald,Lifschitz,Wagner}) process takes place, and the dispersed phase diffuses through the continuous one to reduce the total area of the interface. To quantify the role of the Ostwald ripening in our system, we have simulated a collection of droplets with an initial heterogeneous distribution of radii (ranging in the interval $R \in [10,50]$ lbu) in a computational domain of size $L_x \times L_z= L \times L $ covered by $N_x \times N_z =1024 \times 1024$ lattice sites with full periodic boundary conditions. Similarly to Fig. \ref{fig:disjoining}, two simulations are performed: $i)$ with phase separating interactions \eqref{Phase} only, having strength parameter ${\cal G}_{AB}=0.586$ lbu;  $ii)$ with both phase separating interactions \eqref{Phase} and competing interactions \eqref{NNandNNN} having strength parameters ${\cal G}_{AB}=0.586$ lbu, ${\cal G}_{AA,1}={\cal G}_{BB,1}=-7.0$ lbu and ${\cal G}_{AA,2}={\cal G}_{BB,2}=6.1$ lbu. For fixed space resolution, these parameters correspond to the onset of a stable interface, with a peak in $\Pi(h)$ roughly a $10$\% lower than the one reported in Fig. \ref{fig:disjoining}. In the case of phase separating interactions only, as we can readily appreciate from the snapshots of a sub-region of the system (Panels (a)-(c) of Fig. \ref{fig:coarsening-dilute}), coalescence contributes to the process of interface minimization. Also, smaller domains shrink and finally disappear. The process is more quantitatively analyzed in the bottom panel of Fig. \ref{fig:coarsening-dilute}, where we monitor the average radius $R(t)$ of the droplets as a function of time. On average, a scaling law $R^3(t) \sim t$ is observed, confirming the correctness of the LSW scaling. The tiny jumps witness the subtle role of coalescence, which takes place on time-scales much smaller than the diffusive time-scales associated with the Ostwald ripening. By switching-on the competing interactions \eqref{NNandNNN}, coalescence is inhibited. To highlight this effect, we have chosen a configuration with increased fraction of the dispersed phase (Panels (d)-(f)). Even if coalescence is switched-off, the dispersed phase still diffuses through the continuous phase showing scaling properties compatible with a LSW process, the only difference being a much smaller prefactor.


\begin{figure}[t!]
\subfigure[\,\, t=t$_0$]{\includegraphics[width=0.155\textwidth]{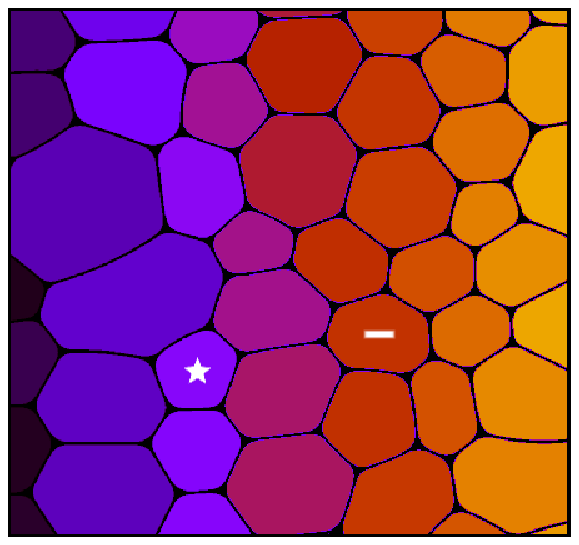}}
\subfigure[\,\, t=t$_0$+$20 \Delta$]{\includegraphics[width=0.155\textwidth]{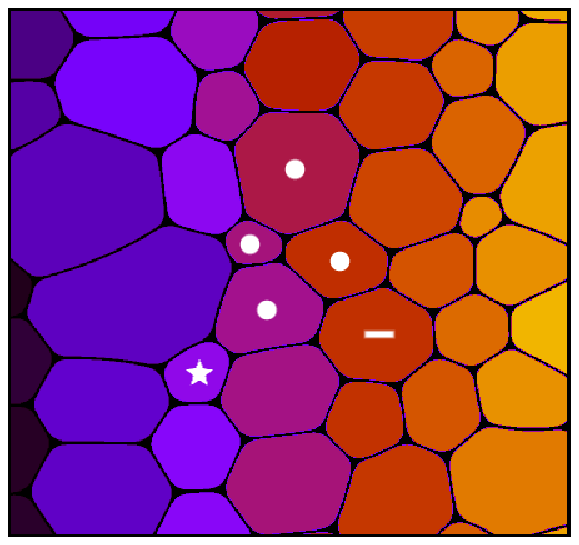}}
\subfigure[\,\, t=t$_0$+$23 \Delta$]{\includegraphics[width=0.155\textwidth]{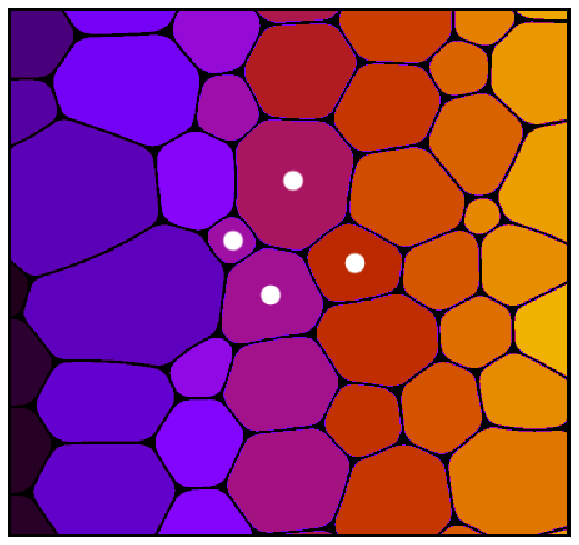}}\\
\includegraphics[width=0.48\textwidth]{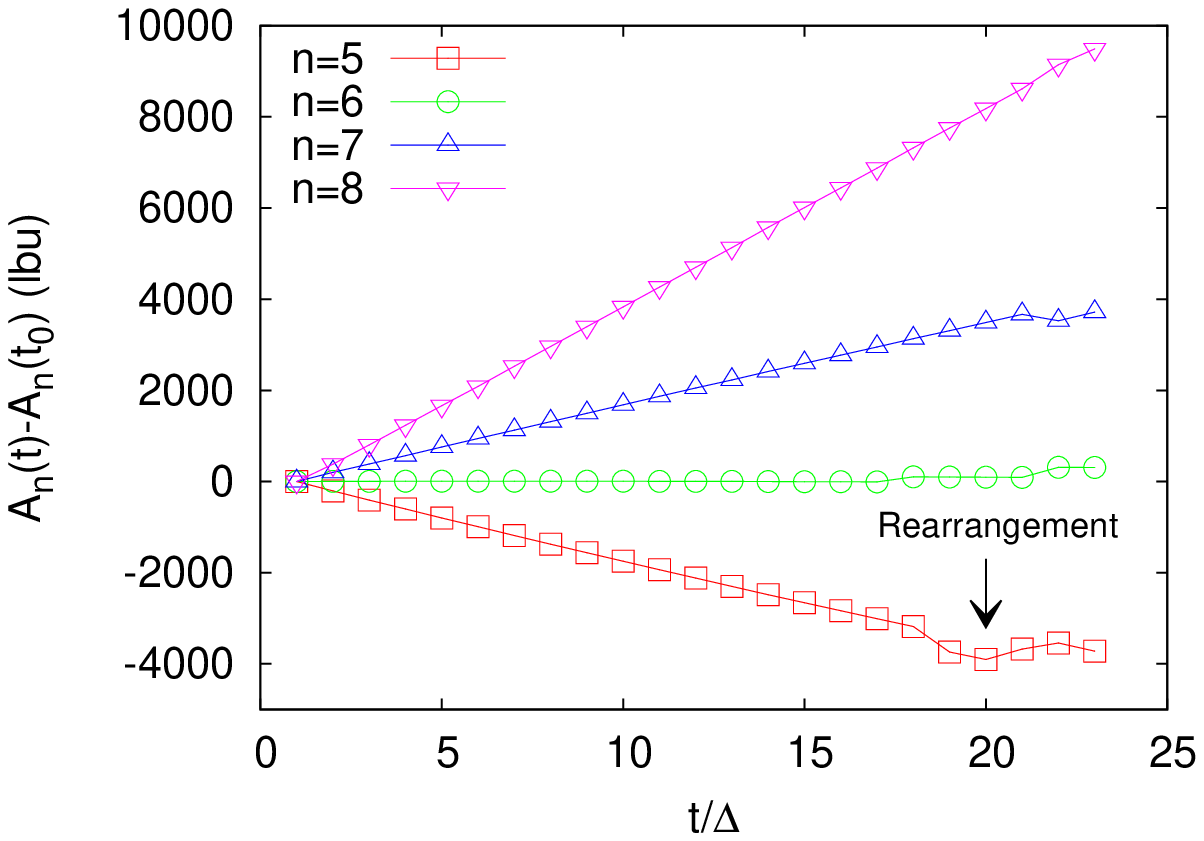}
\caption{Coarsening dynamics in the {\it dry} limit. Panels (a)-(c): Images of a sub-region of the {\it jammed} diphasic system, with different colors referring to different droplets of the dispersed phase. Droplets shrink or grow, depending on their number of sides: droplets with a number of sides larger than six grow (${\bm -}$, Panels (a)-(b)) at the expenses of droplets with a number of sides smaller than six ($\star$, Panels (a)-(b)), whereas six-sided droplets neither grow nor shrink. This is quantitatively checked in the bottom panel, where we monitor the average area $A_n$ of droplets with $n$ sides. Remarkably, the shape of the droplet, the length of its edges, and its set of neighbors do not matter. The various lines indeed represent a fit to Von-Neumann law $dA_n(t)/dt = K(n - 6)$, where $K$ is uniquely adjusted to fit all the data. During the coarsening dynamics T1 rearrangements occur. The four droplets involved in a T1 rearrangement are indicated ($\bullet$, Panels (b)-(c)). $\Delta=50 \times 10^4$ lbu is a characteristic time lapse.}
\label{fig:coarsening-dry}
\end{figure}

With the evidence that the scaling laws for the time dynamics of a dilute system are actually the expected ones \cite{Ostwald,Lifschitz,Wagner,Lambert10}, we finally report in Fig. \ref{fig:coarsening-dry} a quantitative assessment that shows how our model
captures the correct coarsening dynamics of a dry system in a jammed state. A collection of closely packed droplets is simulated, again, in a computational domain of size $L_x \times L_z= L \times L $ covered by $N_x \times N_z =1024 \times 1024$ lattice sites with full periodic boundary conditions. The packing fraction of the dispersed phase in the continuum phase is kept the same and approximately equal to $90 \%$, thus domains of the dispersed phase resemble closely packed polygons (Panels (a)-(b) in Fig. \ref{fig:coarsening-dry}). The degree of polydispersity is defined as the standard deviation of the droplets area, $\sigma(A)$, expressed in units of its mean, i.e. $d=\sigma(A)/\langle A\rangle=0.3$ in our case. For an ideal dry diphasic system of such kind, and under the approximation that the intersection angles of the interfaces exactly satisfy local force balance, individual droplets areas evolve according to Von-Neumann's law \cite{VonNeumann}
\be\label{eq:VonNeumann}
\frac{d A_n(t)}{d t} = K(n - 6)
\ee
where $A_n$ is the average area of droplets with $n$ sides and $K$ is the coefficient of diffusion between adjacent droplets. Figure \ref{fig:coarsening-dry} provides evidence that Eq. \eqref{eq:VonNeumann} is verified. From a visual inspection of the images of a sub-region of the diphasic system during the coarsening dynamics, we actually see that droplets with fewer than six sides shrink (indicated with a $\star$ in Panels (a)-(b)), whereas those with more than six sides grow (indicated with a ${\bm -}$ in Panels (a)-(b)). To calculate the area of the closely packed droplets, we perform a Voronoi tessellation (using the {\it voro++} libraries \citep{Rycroft06}) constructed from the centers of mass of the droplets. When following in time the average area of the droplets, this fits the theoretical prediction \eqref{eq:VonNeumann} quite well, with a diffusion coefficient $K$ which is the same for all $n$'s.\\
Due to the internal coarsening dynamics, also plastic rearrangements in the form of T1 events are present. This is evident from Panels (b)-(c) of figure \ref{fig:coarsening-dry} ($\bullet$ indicates the involved droplets). Plastic rearrangements are also responsible for deviations from the linear Von-Neumann law \eqref{eq:VonNeumann}, indicated with an arrow in the bottom panel of Fig. \ref{fig:coarsening-dry}.\\

\section{Response Under External Oscillatory Strains}\label{sec:response}

\begin{figure}[t!]
\subfigure[\,\,]{\includegraphics[width=0.52\textwidth]{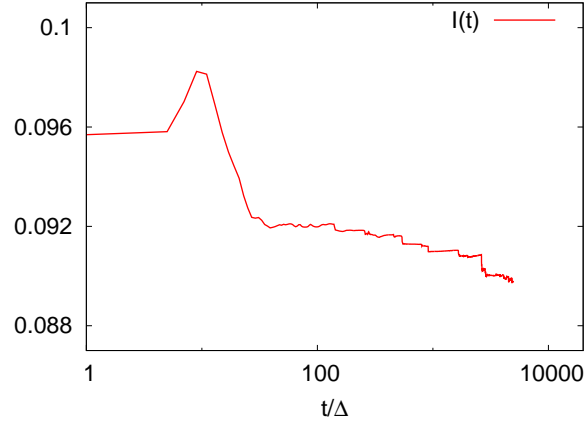}}
\subfigure[\,\,]{\includegraphics[width=0.52\textwidth]{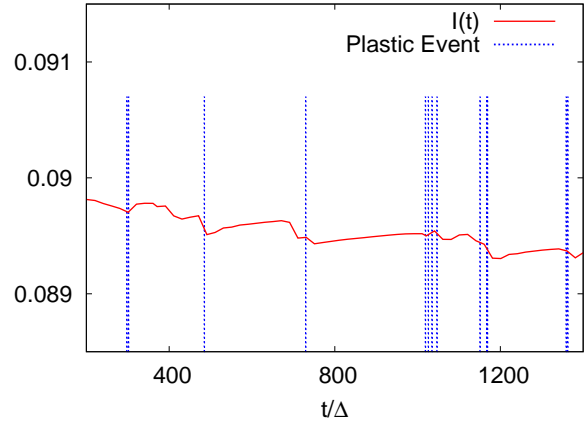}}
\caption{Statistics of plastic rearrangements in the coarsening dynamics. Panel (a): behaviour of the interface length indicator $I(t)$ (see Eq. \eqref{II}) as a function of time. Time is made dimensionless with respect to the time lapse $\Delta=10^3$ lbu. Panel (b): magnified view of panel (a) for $t/\Delta \in [200-1400]$. The vertical spikes indicate when plastic events occur. Note that there is a clear relationship between plastic events and the decrease of $I(t)$. \label{fig:internal-noise}}
\end{figure}

\begin{figure}[t!]
\includegraphics[width=0.46\textwidth]{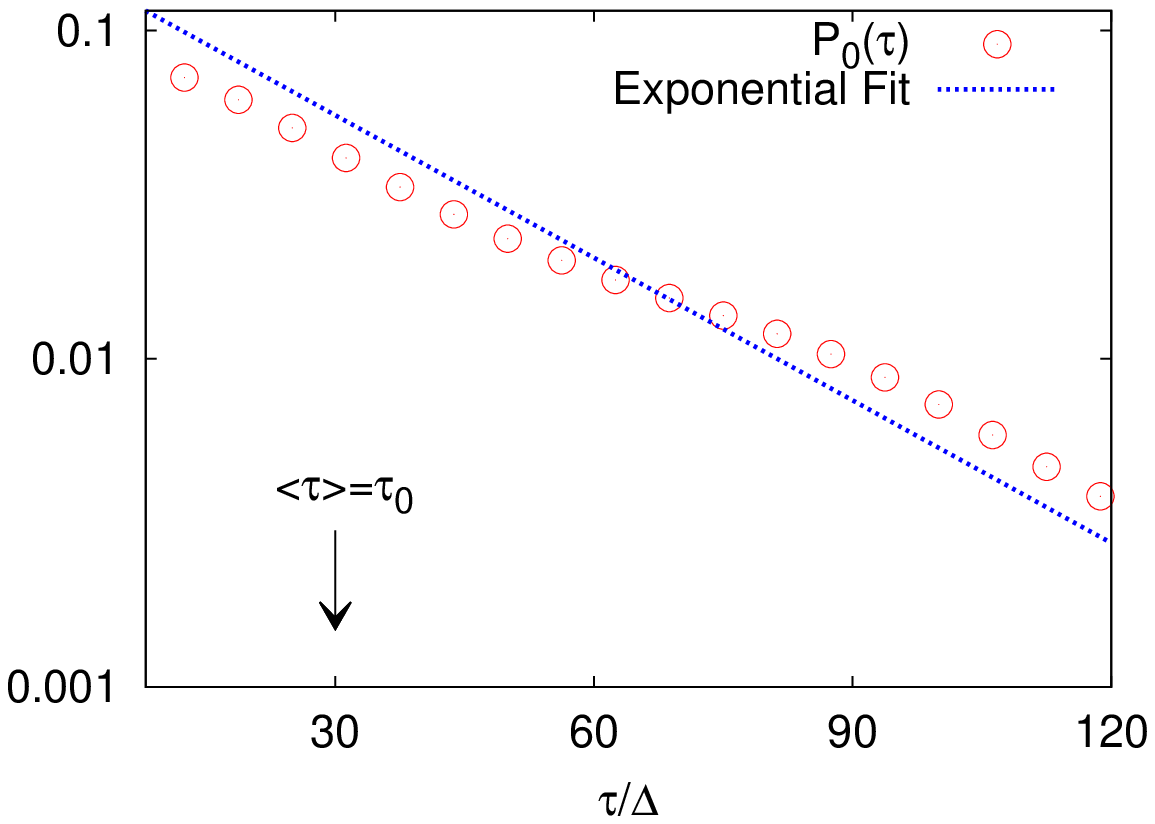}
\caption{Probability density function $P_{0}(\tau)$ of the time interval $\tau$ between two successive plastic events (the subscript $0$ indicates unstrained conditions). $P_0(\tau)$ is close to an exponential distribution with an average value of $\langle \tau \rangle_0=\int P_0(\tau) \tau d \tau = \tau_0 \approx 3 \times 10^4$ lbu (indicated with an arrow). Time is made dimensionless with respect to the time lapse $\Delta=10^3$ lbu. \label{fig:internal-noise-durian}}
\end{figure}

The basic question we wish to address in this section is whether plastic rearrangements can be regarded as activated processes induced by a suitable ``noise'' produced by the internal dynamics. This requires a systematic characterization of plastic events and their statistical occurrences as a function of an external strain. To perform long-time simulations with an acceptable computational effort, we can decrease the resolution by using a computational domain of size $L_x \times L_z= L \times L $ covered by $N_x \times N_z =512 \times 512$ lattice sites. In order to keep the same packing fraction and the same number of droplets, we must increase the disjoining pressure at the interface: a proper parameters choice to that purpose is ${\cal G}_{AB}=0.586$ lbu, ${\cal G}_{AA,1}={\cal G}_{BB,1}=-9.0$ lbu and ${\cal G}_{AA,2}={\cal G}_{BB,2}=8.1$ lbu, which increases the peak in $\Pi(h)$ shown in Fig. \ref{fig:disjoining} by roughly a factor 2. Moreover, we confine the system between two parallel walls with no-slip boundary conditions. For unstrained situations (i.e. no external load), the coarsening dynamics of the system is illustrated and characterized in Fig. \ref{fig:internal-noise}. Panel (a) reports the time behaviour of the following interface length indicator:
\begin{equation}\label{II}
I(t) = \frac{1}{L^2} \int |{\bm \nabla} \phi ({\bm r})|^2 d {\bm r}
\end{equation}
where $\phi=\rho_A-\rho_B$ is the density difference. In Panel (b) of Fig. \ref{fig:internal-noise} we show the behaviour of $I(t)$ during a period of $1.2 \times 10^6$ lbu, with the indication of the plastic events observed during the evolution (vertical spikes). From these observations, we realize that, because of spatial disorder, rearrangements occur at random instants and random space locations. To characterize the distribution of rest times between successive events \cite{Durian08,FOAM}, we consider the quantity $\tau_i \equiv t_{i}-t_{i-1}$, where $t_i$ is the time at which the $i$-th plastic event occurs. In Fig. \ref{fig:internal-noise-durian} we show the probability density function (pdf) $P_0(\tau)$ of $\tau$ (the subscript $0$ indicates unstrained conditions) that fits reasonably well an exponential distribution. The internal characteristic time-scale of the system (i.e. the average of $\tau$) is estimated to be $\langle \tau \rangle_0=\int P_0(\tau) \tau d \tau = \tau_0 \approx 3 \times 10^4$ lbu. It is noticeable that the distribution $P_0(\tau)$ bears close similarities with the distribution of rest times for bubbles rearrangements driven by coarsening in real aqueous foams \cite{Durian08}. Moreover, numerical simulations based on the surface evolver method  have shown that the temporal statistics of coarsening-induced bubble rearrangements  can be described by a Poisson process \cite{FOAM}.


\begin{figure}[t!]
\begin{center}
\includegraphics[width=0.6\textwidth]{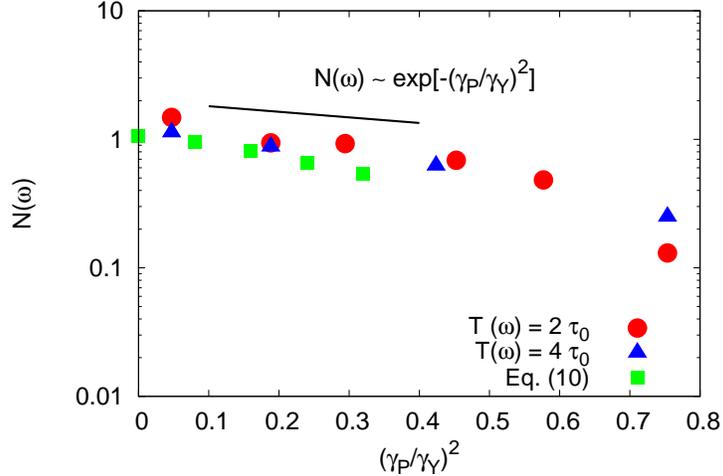}
\end{center}
\caption{Behaviour of $N(\omega)=\langle \tau \rangle_{\omega}/T(\omega)$ as a function of $(\gamma_P/\gamma_Y)^2$, where $T(\omega)$ is the period of the external strain forcing, $\langle \tau \rangle_{\omega}$ is the average time observed between two consecutive plastic events (the subscript $\omega$ refers to strained conditions), $\gamma_P$ is the peak value of the external strain applied to the system and $\gamma_Y$ is the yield strain. Two values of $T(\omega)$ are considered, $T(\omega)= 2\, \tau_0 $ and $T(\omega)=4 \, \tau_0$, with $\tau_0$ the internal characteristic time scale, taken as the average time between successive plastic events in the coarsening dynamics (unstrained conditions, see Fig. \ref{fig:internal-noise}). In both cases, the ratio $\langle \tau \rangle_{\omega}/T(\omega)$ decays exponentially in $(\gamma_P/\gamma_Y)^2$ (the bar is a guide for the eyes). The green squares correspond to $N(\omega)$ found in the naive double well potential equation \eqref{SR}. \label{fig:activated-process-hint}}
\end{figure}

To gain further insight, we now consider the case of an external oscillatory strain
\be
\gamma(t)=\gamma_P \sin (\omega t)
\ee
with frequency $\omega$ and peak amplitude $\gamma_P$ smaller than the yield strain $\gamma_Y$. By using data at very low frequency \cite{EPL13}, we obtain a yield strain $\gamma_Y \sim 0.3$, a yield stress $\sigma_Y \sim 1.1 \times 10^{-4}$ lbu and an elastic load modulus $E=4.4 \times 10^{-4} \rho$ lbu, where $\rho=1.32$ lbu is the total density. For small $\gamma_P$, we can assume that the external strain provides an energy input in the system $\sim E \gamma_P^2/2$,  to be compared with the energy barrier for a plastic event, which can be estimated $\sim E\gamma_Y^2/2$. In Fig. \ref{fig:activated-process-hint},  we report the number $N(\omega)=\langle \tau \rangle_{\omega} / T(\omega)$ defined as the ratio of the average time $\langle \tau \rangle_{\omega}$ between two consecutive plastic events (the subscript $\omega$ indicates strained conditions) and the forcing period $T(\omega) \equiv 2 \pi/\omega$. Note that, by definition, $1/N(\omega)$ is just the number of plastic events per cycle. In the figure, we plot $N(\omega)$ in log scale versus $(\gamma_P/\gamma_Y)^2$ for two oscillatory strains with period $T(\omega)= 4\,\tau_0 $ (blue triangles) and $T(\omega)= 2 \,\tau_0 $ (red bullets). Since $(\gamma_P/\gamma_Y)^2$ represents the ratio of the energy induced by the external strain with respect to the energy barrier, we expect $\log(N(\omega))$ to decrease linearly with $(\gamma_P/\gamma_Y)^2$ in an activated process. This is exactly what Fig. \ref{fig:activated-process-hint} shows for both $T(\omega)=4 \tau_0$ and $T(\omega)=2 \tau_0$. Figs. \ref{fig:internal-noise} and \ref{fig:activated-process-hint} support the idea that the internal (deterministic) dynamics acts like an intrinsic noise in the system.  If this idea of noise-induced activated processes is correct, it is then natural to look for a connection between plastic events and a general mechanism like {\it stochastic resonance} (SR) \cite{SR1,SR2}.

\subsection{Stochastic Resonance}

\begin{figure}[t!]
\begin{center}
\includegraphics[width=0.6\textwidth]{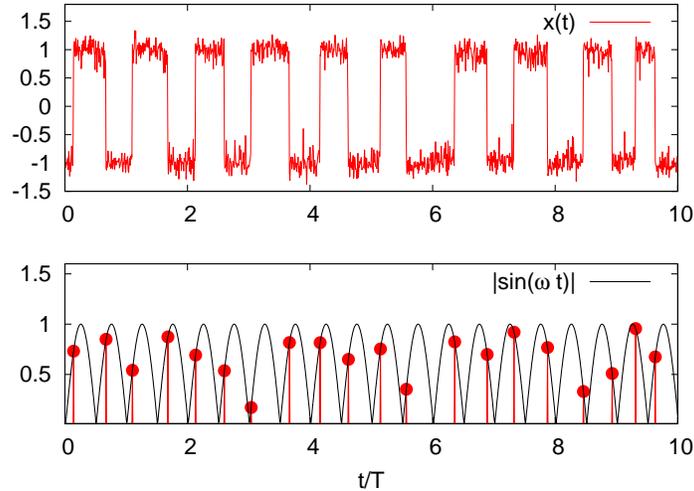}
\end{center}
\caption{Typical signature of stochastic resonance for the system described by Eq. \eqref{SR} with $\Delta V = 1/4$, $\epsilon = 0.0625$, $A = 0.08$. The period $T = 2\pi/\omega = 12 \times 10^3$ of the applied forcing is twice the internal time scale of the system $\tau_{x,0}=\frac{\pi}{\sqrt{2}} \mbox{exp}(2 \Delta V/\epsilon) \approx 6 \times 10^3$. The system shows a nearly periodic transition between the two minima (upper panel). In the lower panel we plot the strength of the external forcing $|\sin(\omega t)|$ (the amplitude has been set equal to 1 for simplicity), whereas the red bullets indicate the transition from one state to the other. In this conceptual picture, the red bullets are equivalent to the plastic events occurring in the dynamics of the soft-glassy system.}
\label{fig:SR}
\end{figure}

We begin by shortly reviewing the main distinctive features of SR. The mechanism of stochastic resonance can be explained by considering a sample stochastic differential equation
\begin{equation}\label{SR}
\frac{dx(t)}{dt} = x(1-x^2)+A \sin(\omega t) + \sqrt{\epsilon}\, W(t)
\end{equation}
representing the Lagrangian trajectory $x(t)$ of a point with unitary mass under the action of the external potential $V(x)=(1-x^2)^2/4$ and forcing term $A \sin(\omega t)$ with adjustable strength parameter $A$ and frequency $\omega$. The term $W(t)$ is a white noise delta-correlated in time with intensity $\epsilon$, i.e. $\langle W(t) W(t^{\prime})\rangle = \epsilon \delta(t-t^{\prime})$.  When $A=0$ (zero external forcing), the system shows transitions between the two minima of the double-well potential, exponentially distributed in time with an average $\tau_{x,0}=\frac{\pi}{\sqrt{2}} \mbox{exp}(2 \Delta V/\epsilon)$ \cite{SR1,SR2}, where, for the present case, the potential barrier is $\Delta V = 1/4$. The effect of $A \ne 0$ induces a variation in the potential barrier, $\Delta V = 1/4 + A \sin(\omega t)$. In Fig. \ref{fig:SR}, we show the classic signature of SR: noise induced hopping between the potential wells can become synchronized with the periodic forcing. In particular, in the bottom panel of Fig. \ref{fig:SR} we show the strength of the external forcing (the amplitude has been set equal to 1 for simplicity), while the bullets report the times at which jumps between the two wells occur, highlightning the syncronization effect. This statistical synchronization takes place when the average rest time $\tau_{x,0}$ between two noise-induced transitions is comparable with half of the period of the external forcing, leading to the well-known {\it matching condition} for SR\cite{SR1,SR2}
\be\label{eq:matching}
T(\omega)=\frac{2 \pi}{\omega} \approx 2 \tau_{x,0}.
\ee
SR is a pretty general and fairly robust mechanism; it has been observed for both stochastic and deterministic chaotic systems (where the chaotic dynamics plays the role of the noise) with and without the existence of multiple metastable states.



\subsection{Connection with LB numerical simulations}

\begin{figure}[t!]
\begin{center}
\includegraphics[width=0.6\textwidth]{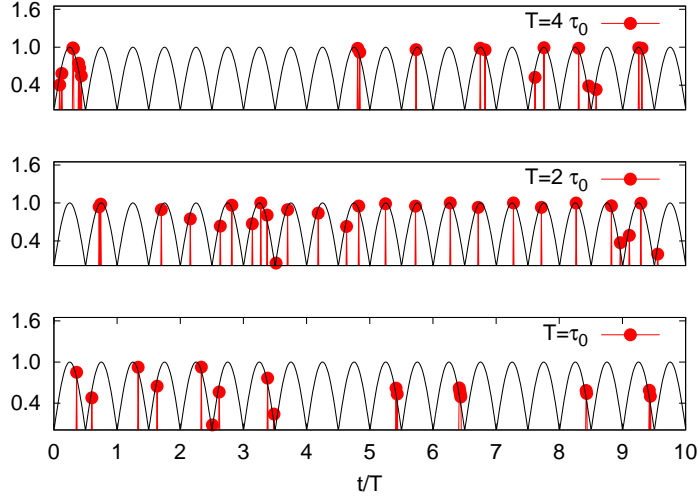}
\end{center}
\caption{Plastic events observed during $10$ cycles for the strain $\gamma(t)=\gamma_P \sin (\omega t)$ with $\gamma_P /\gamma_Y=0.4$ for three different periods: $T(\omega)=4\,\tau_0$ (upper panel), $T(\omega)= 2\,\tau_0$ (middle panel) and $T(\omega)= \tau_0$ (lower panel). The characteristic time $\tau_0$ represents an internal time-scale, taken as the average time between successive plastic events in the coarsening dynamics (see Fig. \ref{fig:internal-noise}). The solid line represents $|\sin(\omega t)|$. For the period $T(\omega)= 2\, \tau_0$ we observe one of the hallmarks of stochastic resonance, i.e. plastic events are in phase with the applied strain. \label{fig:SR-LB}}
\end{figure}

\begin{figure}[t!]
\begin{center}
\includegraphics[width=0.5\textwidth]{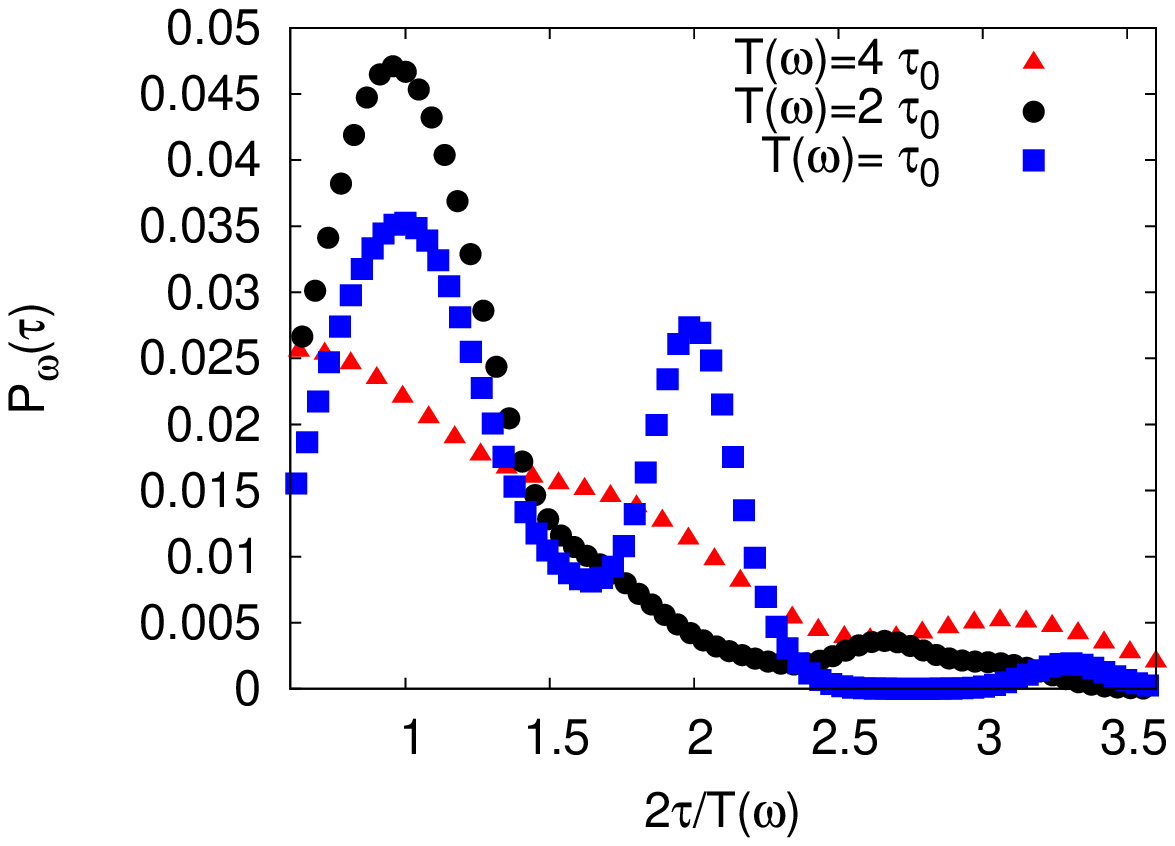}
\includegraphics[width=0.5\textwidth]{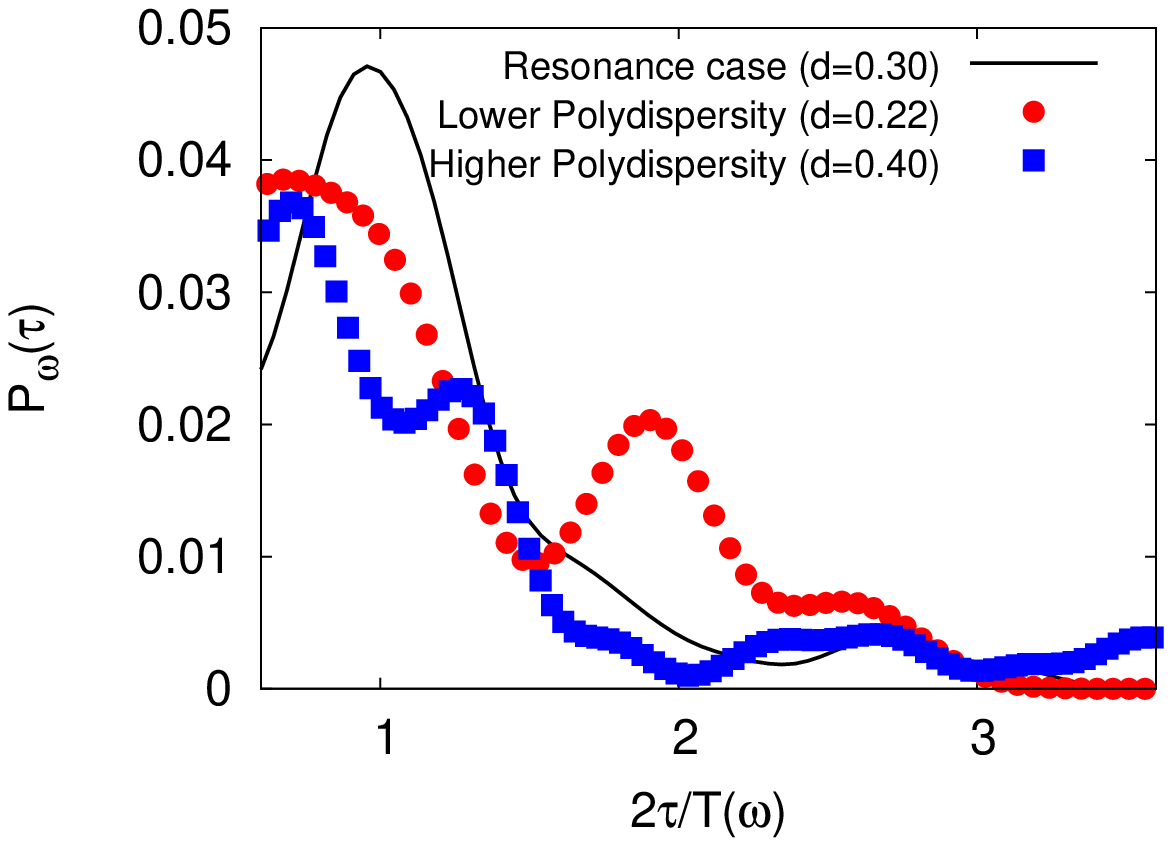}
\end{center}
\caption{Top Panel: probability density function $P_{\omega}(\tau)$ of the time interval $\tau$ between two successive plastic events at different periods with a fixed degree of polydispersity, $d=0.22$ (see text for details). $P_{\omega}(\tau)$ is plotted as a function of $2\,\tau/T(\omega)$ and the signature of stochastic resonance is a narrow peak for $\tau = 0.5\,T(\omega)$ when $T(\omega)= 2\, \tau_0$. Note that for $T(\omega)= \tau_0$, $P_{\omega}(\tau)$ shows two clearly defined peaks at $\tau = 0.5\,T(\omega)$ and $\tau=T(\omega)$. Bottom panel: probability density function of $\tau$ for lower and higher polydispersity with respect to the case $T(\omega)=2 \tau_0$ (labeled as ``resonance case'') already reported in the top panel. We speculate that the noise is a direct result of the system polydispersity: the narrow peak at $\tau=0.5\,T(\omega)$ observed in the resonant case disappears when changing the degree of polydispersity $d$. \label{fig:LB-probability-SR}}
\end{figure}

We seek signatures of SR, in our simulated system, by changing the period $T(\omega)$ of the external oscillatory strain. Although a formal link between our simulated system and the prototypical SR model equation discussed in the previous section remains to be established, the results shown in Fig. \ref{fig:activated-process-hint} support nonetheless the idea  that plastic rearrangements can be regarded as activated processes induced by a suitable noise effect.  If the source of noise is provided by the internal coarsening dynamics described in Section \ref{sec:aging}, an external oscillatory strain with period $T(\omega) \approx 2 \tau_0$ should exist, for which  syncronization between the external strain and plastic events could be observed. In other words, SR could be detected in two different ways: $i)$ plastic events should be in phase with the strain and $ii)$ the probability density function of $\tau$ should be peaked around $\tau=0.5\,T(\omega)$. We performed simulations with the external periodic strain for three different periods $4\,\tau_0$, $2\,\tau_0$ and $\tau_0$. The external strain amplitude is such that $\gamma_P/\gamma_Y=0.4$. To draw a parallel with the double well potential of Eq. \eqref{SR}, our choice $\gamma_P/\gamma_Y=0.4$ corresponds to $A=0.04$, which is the smallest amplitude at which SR is observed. Moreover, $\gamma_P$ is supposed to be the smallest amplitude in the external strain for which SR can be observed in the system. In Fig. \ref{fig:SR-LB}, we show a snapshot of $10$ cycles for the three different periods. The solid line corresponds to $|\sin(\omega t)|$, whereas the red bullets indicate the occurrence of plastic events. In the top panel of Fig. \ref{fig:LB-probability-SR}, we show the pdf of $\tau$, as obtained from $75$ cycles of the external forcing. Both figures clearly show that SR occurs with the right signature: plastic events are nearly in phase with the forcing and the pdf of $\tau$ is peaked around $\tau=0.5\,T(\omega)$. Note that at $T(\omega) = \tau_0$ the pdf shows two clear peaks at $\tau=0.5 \, T(\omega)$ and $\tau= T(\omega)$. Actually, when $T(\omega) = \tau_0$, we may qualitatively think of our system as a collection of different spatial regions where the probability that a plastic event occurs reaches a maximum at the largest strain (in absolute value), i.e. at times $T_n = (n+1)T (\omega)/2$ ($n=0,1,...$). Thus we expect that the probability density function $P_{\omega}(\tau)$ should show well defined peaks at $\tau=T_n$, in agreement with the results shown in Figs. \ref{fig:SR-LB} and \ref{fig:LB-probability-SR}. Finally, it is interesting to observe that the green squares shown in Fig. \ref{fig:activated-process-hint} correspond to $N(\omega)$ found in the naive double well potential equation \eqref{SR}, with $\omega = 2\pi/\tau_{x,0}$, i.e., at the SR frequency. As a matter of fact, the slope of $\log{(N(\omega))}$ is close to the one observed in our simulations, lending further support to the idea that the internal dynamics can be regarded as a form of intrinsic noise.


As we discussed in Section \ref{sec:aging}, space randomness (due to different sizes and shapes of droplets) is underlying the coarsening dynamics. We argue, therefore, that increasing or decreasing space randomness (i.e. changing polydispersity) is equivalent to change the internal noise. If this is true, the narrow peak at $\tau=0.5 \, T(\omega)$, upon increasing or decreasing of the noise amplitude away from the resonance value, should disappear. This is indeed observed in our case: we repeated the simulation at the resonance frequency either by increasing or by decreasing the polydispersity of the system with all the other parameters being kept at the same value. With no external strain, the number of plastic events decreases (increases) by decreasing (increasing) the polydispersity by a factor $\approx 30\%$. In the bottom panel of Fig. \ref{fig:LB-probability-SR} we show the probability density function $P_{\omega}(\tau)$ for both cases of ``weak'' (lower polydispersity, $d=0.22$) and ``strong'' (higher polydispersity, $d=0.4$) noise in presence of a periodic strain. It is clear that upon increasing or decreasing the spatial disorder, SR fades away in accordance with the results expected from Eqs. \eqref{SR}-\eqref{eq:matching}. We hasten to add that this interpretation of the noise as a direct result of the system polydispersity, although pretty plausible, is by no means unique and surely in need of consolidation through further systematic investigations.


\section{Summary and Outlook}\label{sec:conclusions}


Summarizing, our numerical simulations provide evidence that plastic rearrangements can be likened to activated processes, induced by a suitable form of intrinsic ``noise''. In particular, the idea of coarsening-induced noise is here supported by the analysis of plastic events under time oscillating strains, where it is observed that the number of plastic rearrangements depends exponentially on the external energy-supply. We have also studied the interplay between the internal characteristic time-scale of the coarsening dynamics and the external time-scale associated with the imposed oscillating strain, showing that for suitably chosen frequencies, the system exhibits the phenomenon of stochastic resonance (SR). Taken all together, these points lend further credit to several results and conjectures presented in the soft-glassy literature, particularly to the central idea of effective noise-induced activated escape from free energy random traps \cite{SGR1,SGR2}. However, we wish to highlight that the concept of an effective noise in soft glasses was invoked to explain the linear response in the rheological properties. In contrast, the work on SR presented here belongs to the non-linear regime of finite amplitude oscillations, the amplitude being a finite fraction of the yield strain $\gamma_Y$.  Indeed, we find that there is a minimal amplitude of $\gamma_P/\gamma_Y$  (see Fig. \ref{fig:SR-LB}) below which SR is not detected. \\
An interesting point of discussion emerges from the attempt of connecting the present results with experimental data on soft-glasses. The (exponential) distribution of rest times in unstrained conditions (see Figs. \ref{fig:internal-noise} and \ref{fig:internal-noise-durian}) is very similar to the probability density function for the rest time between successive bubble rearrangements in real aqueous foams \cite{Durian08,FOAM}. Thus, we expect that the results presented in this paper could have an experimental confirmation. We remark that in actual experiments the rate of the plastic events is expected to slow down with increasing foam age. However, for the characteristic timescales given in the literature \cite{Durian91,Addad01}, it seems possible to achieve the SR matching conditions in a window of time shorter than the rate of change of the internal noise. \\

M. Sbragaglia, A. Scagliarini \& R. Benzi kindly acknowledge funding from the European Research Council under the European Community's Seventh Framework Programme (FP7/2007-2013)/ERC Grant Agreement No. 279004. P. Perlekar \& F. Toschi acknowledge partial support from the Foundation for Fundamental Research on Matter (FOM), which is part of the Netherlands Organisation for Scientific Research (NWO).


\bibliography{rsc}       
\bibliographystyle{rsc} 
\end{document}